\documentclass[entropy,article,accept,pdftex,moreauthors]{Definitions/mdpi}

\firstpage{1} 
\pubvolume{1}
\issuenum{1}
\articlenumber{0}
\pubyear{2024}
\copyrightyear{2023}
\externaleditor{Academic Editor: Paolo Castiglioni}
\datereceived{20 November 2023} 
\daterevised{23 December 2023} 
\dateaccepted{26 December 2023} 
\datepublished{ } 
\hreflink{https://doi.org/} 

\usepackage[all]{nowidow}
\graphicspath{{figures/}}
\newcolumntype{L}[1]{>{\raggedright\arraybackslash}m{#1}}

\Title{\highlighting{Novel} 
 Entropy-Based Metrics for Long-Term Atrial Fibrillation Recurrence Prediction Following Surgical Ablation: Insights from Preoperative Electrocardiographic Analysis}

\TitleCitation{Novel Entropy-Based Metrics for Long-Term Atrial Fibrillation Recurrence Prediction Following Surgical Ablation: Insights from Preoperative Electrocardiographic Analysis}

\Author{\hl{Pilar} 
 Escribano $^{1}$\orcidA{}, Juan R\'odenas $^{1}$\orcidB{}, Manuel Garc\'ia $^{1}$\orcidC{}, Fernando Hornero $^{2}$\orcidD{}, Juan M. Gracia-Baena $^{2}$\orcidE{},\linebreak Ra\'ul Alcaraz $^{1}$\orcidF{} and Jos\'e J. Rieta $^{3,}$*\orcidG{}}

\AuthorNames{Pilar Escribano, Juan R\'odenas, Manuel Garc\'ia, Fernando Hornero, Juan M. Gracia-Baena, Ra\'ul Alcaraz and Jos\'e J. Rieta}

\AuthorCitation{\hl{Escribano,} 
 P.; R\'odenas, J.; Garc\'ia, M.; Hornero, F.; Gracia-Baena, J.M.; Alcaraz, R.; Rieta, J.J.}

\address{%
$^{1}$ \quad Research Group in Electronic, Biomedical and Telecommunication Engineering, University of Castilla-La Mancha, 02071 Albacete, Spain; pilar.escribano@uclm.es (P.E.);  juan.rodenas@uclm.es (J.R.); manuel.garcia@uclm.es (M.G.); raul.alcaraz@uclm.es (R.A.)
\\
$^{2}$ \quad Cardiovascular Surgery Department, Hospital Cl\'inico Universitario de Valencia, 46010 Valencia, Spain; hornero\_fer@gva.es (F.H.); gracia\_juabae@gva.es (J.M.G.-B.)
\\
$^{3}$ \quad BioMIT.org, Electronic Engineering Department, Universitat Politecnica de Valencia, 46022 Valencia, Spain}

\corres{Correspondence: jjrieta@upv.es}

\abstract{Atrial fibrillation (AF) is a prevalent cardiac arrhythmia often treated concomitantly with other cardiac interventions through the Cox--Maze procedure. This highly invasive intervention is still linked to a long-term recurrence rate of approximately 35\% in permanent AF patients. The aim of this study is to preoperatively predict long-term AF recurrence post-surgery through the analysis of atrial activity (AA) organization from non-invasive electrocardiographic (ECG) recordings. A dataset comprising ECGs from 53 patients with permanent AF who had undergone Cox--Maze concomitant surgery was analyzed. The AA was extracted from the lead V1 of these recordings and then characterized using novel predictors, such as the mean and standard deviation of the relative wavelet energy ($RWEm$ and $RWEs$) across different scales, and an entropy-based metric that computes the stationary wavelet entropy variability ($SWEnV$). The individual predictors exhibited limited predictive capabilities to anticipate the outcome of the procedure, with the $SWEnV$ yielding a classification accuracy (Acc) of 68.07\%. However, the assessment of the $RWEs$ for the seventh scale ($RWEs7$), which encompassed frequencies associated with the AA, stood out as the most promising individual predictor, with sensitivity (Se) and specificity (Sp) values of 80.83\% and 67.09\%, respectively, and an Acc of almost 75\%. Diverse multivariate decision tree-based models were constructed for prediction, giving priority to simplicity in the interpretation of the forecasting methodology. In fact, the combination of the $SWEnV$ and $RWEs7$ consistently outperformed the individual predictors and excelled in predicting post-surgery outcomes one year after the Cox--Maze procedure, with Se, Sp, and Acc values of approximately 80\%, thus surpassing the results of previous studies based on anatomical predictors associated with atrial function or clinical data. These findings emphasize the crucial role of preoperative patient-specific ECG signal analysis in tailoring post-surgical care, enhancing clinical decision making, and improving long-term clinical outcomes.}


\newcommand{\textmk}[1]{\textcolor{black}{#1}}

\begin{document}


\section{Introduction}

Atrial fibrillation (AF) is one of the most prevalent supraventricular rhythm disorders, impacting about 0.51\% of the world’s population and affecting more than 37.5~million individuals~\cite{Lippi:2021aa}. The likelihood of suffering from this arrhythmia increases with age; only 0.16\% of people under 50 have this disorder, whereas this proportion increases to 17\% in people over 80 \cite{zoni2013epidemiology}. The expected increase in the population aged over 60 from 962 million in 2017 to more than 2 billion in 2050~\cite{unitednations2017world}, the higher occurrence of chronic conditions that make older people more prone to AF, and the latest advancements in the detection of arrhythmia are expected to lead to an increase in the incidence of this cardiac pathology~\cite{hindricks2021}. In 2016, the estimated number of European Union older persons affected by arrhythmia was approximately 3.18, 1.72, and 2.71 million cases for paroxysmal, persistent, and permanent AF. By 2060, these numbers are expected to increase to approximately 5.99, 2.83, and 5.60 million cases, respectively, solidifying AF as a global epidemic~\cite{Carlo2020}. It is estimated that AF management accounts for roughly 1\% of the budget for most health services in developed countries~\cite{morin2016state}. A more recent study highlighted the significantly higher burden of medical visits among patients with AF, resulting in a mean total healthcare cost that was USD 27,896 higher than that of non-AF patients~\cite{deshmukh2022}. This situation carries important economic implications for every health system, becoming one of the main challenges in public health~\cite{zoni2013epidemiology}.

For the management of AF, oral anticoagulants are prescribed in the early stages to prevent strokes~\cite{morin2016state}, and all patients receive treatment with a rate control strategy to prevent heart failure and tachycardiomyopathies~\cite{hindricks2021}. However, combining techniques for rhythm and rate control has not demonstrated benefits over their individual application~\cite{carlsson2003randomized}. Hence, current guidelines recommend individualized treatment decisions based on the probability of maintaining sinus rhythm (SR) in the long term~\cite{hindricks2021}. Thus, for patients with persistent symptoms, even under rate control treatment, SR restoration is essential to improve their quality of life~\cite{morin2016state}. For this purpose, the techniques used today include antiarrhythmic drugs, electrical cardioversion (ECV), catheter ablation (CA), and Maze surgery. They are often combined to enhance their effectiveness, although they still fall short of clinical desirability~\cite{morin2016state,vanwagoner2015progress}. Therefore, for symptomatic patients with recent onset AF, the initial treatment typically involves antiarrhythmic drugs~\cite{hindricks2021,mont2017atrial}. Although these drugs can inhibit the arrhythmogenic foci that trigger AF, they do not alter the structural characteristics of the atrial substrate and cannot prevent the arrhythmia from recurring in many cases~\cite{kik2017intra}.

Due to this, CA has become a widely used option in recent years for restoring SR in patients who do not respond favorably to drug treatment and experience strong symptoms~\cite{anselmino2016catheter}. Indeed, current AF management guidelines position CA as a better choice for antiarrhythmic drugs for symptom improvement and mid-term SR maintenance~\cite{hindricks2021}. However, many patients may require multiple CA procedures to effectively manage the arrhythmia. In this regard, it is important to note that the success rate for maintaining SR without severe symptomatic recurrence of AF is around 70\% for patients with paroxysmal AF and approximately 50--60\% for those with persistent AF. Hence, alternative rhythm control options should be considered for patients experiencing recurrences~\cite{Schmidt:2020aa,Dretzke2020}.

Among the surgical alternatives available, Cox--Maze surgery stands out as the most efficient long-term treatment. However, it still presents a limited success rate of approximately 65\% within the first year for patients with permanent AF~\cite{hindricks2021,Sharples2018}. The purpose of Cox--Maze surgery is to eliminate possible re-entry circuits using auricular lesions to prevent fibrillatory conduction and create a specific route able to guide sinus node impulses to the atrioventricular node~\cite{cox2014brief}. Since its introduction in 1987 by Dr. James L. Cox, the technique has evolved. The first two versions of this intervention were discarded due to the technical complexity and the required high rate of pacemaker implantation~\cite{damiano2011cox}.

The third version of the procedure, Cox--Maze III, aimed to stop potential re-entry pathways from causing irregular heartbeats through incisions intended to establish an electrical maze on the atrial tissue~\cite{prasad2003cox}. Subsequently, the Cox--Maze IV technique was clinically introduced in 2002 and is regarded as the current gold standard for surgical AF ablation~\cite{damiano2011cox}. It uses cryothermal  and/or radiofrequency ablation lines instead of the ``cut and sew’’ incisions of its previous version, lowering the risks and technical difficulty without affecting the results~\cite{henn2015late}. Cox--Maze surgery is performed either as a standalone procedure or in combination with other heart interventions, including surgery of mitral and/or aortic valves, or coronary artery bypass grafting~\cite{damiano2011cox}. Concomitant surgery often yields better outcomes in terms of AF elimination compared to exclusive AF surgery, without an increase in mortality or morbidity in patients who have undergone this technique, although pacemaker implantation is more common in such cases~\cite{hindricks2021}.

Following Cox--Maze surgery, patients are prescribed antiarrhythmic and anticoagulant medications upon discharge from the hospital, with follow-up evaluations scheduled at 3, 6, and 12 months post-surgery. If stable SR remains after several months, the drugs are gradually discontinued~\cite{engelsgaard2018long}. However, in case of AF recurrence, ECV is employed to restore SR~\cite{damiano2011cox,henn2015late}. Obtaining preoperative insights into the likelihood of AF recurrence subsequent to Cox--Maze surgery serves multiple strategic purposes. Firstly, it empowers healthcare professionals to plan a more vigilant and tailored postoperative follow-up regimen, ensuring timely interventions, if necessary. For patients deemed to have a lower probability of sustaining SR, this predictive information provides a valuable basis for adjusting their treatment strategy. Furthermore, such data can help guide clinical decision making, steering clear of potentially aggressive drug treatments for individuals whose likelihood of SR maintenance is higher, thereby minimizing their associated side effects. Furthermore, preoperatively assessing the likelihood of AF recurrence after Cox--Maze surgery could limit the higher associated treatment costs and unnecessary risks for patients with potentially limited success in Cox--Maze surgery and could lead to a more suitable treatment option for them to enhance their quality of life, among other benefits~\cite{hindricks2021}.

Most previous studies dealing with preoperative predictions of Cox--Maze surgery have presented limited outcomes from the point of view of clinical application, emphasizing clinical and anatomical information associated with atrial function~\cite{chen2004preoperative,wu2017,Cao2017,jiang2023machine}. In particular, Jiang and colleagues~\cite{jiang2023machine} developed multiple machine learning models based on 58 clinical variables, demonstrating that the most significant preoperative features for predicting AF recurrence were the duration of the arrhythmia, left ventricular ejection fraction, left atrial diameter, neutrophil--lymphocyte ratio, and heart rate. Other studies have also extolled the high predictive capacity of some of these metrics, e.g., AF duration and preoperative left atrial diameter, yielding accuracy (Acc) values of about 75\%~\cite{wu2017}, whereas others yielded much lower Acc values of 55.3 and 68.7\%, respectively, and highlighted the performance of other metrics such as the heart rate, with an Acc of 75.8\%~\cite{Cao2017}. Therefore, these contrasting results among studies underscore the lack of robust evidence supporting the consideration of these anatomical features as singularly reliable predictors.
	
In the last few years, several works have proven that the analysis of atrial activity (AA) organization from surface electrocardiograms (ECGs) is a useful approach to predicting AF termination in a variety of scenarios, including ECV~\cite{alcaraz2010review} and paroxysmal AF spontaneous termination~\cite{Chiarugi2007}. In this context, Hernandez et~al.~\cite{hernandez2014preoperative} addressed the only study focusing on the analysis of surface ECGs and signal-processing features to predict the outcome of Cox--Maze surgery. They examined parameters such as the amplitude of the fibrillatory waves ($fWP$) and the organization of AA through features like the dominant atrial frequency ($DAF$) and sample entropy ($SampEn$), achieving promising predictive results, which were validated in a limited database. Although this study was designed to make a prediction at discharge, the proposed features could also have a role in the long-term prediction horizon.

The current study introduces an innovative approach aimed at preoperatively predicting the restoration of SR one year following the Cox--Maze procedure for patients with permanent AF through the analysis of their AA organization from non-invasive ECG recordings. Novel features are introduced to characterize the variability in the morphological pattern of AA in AF rhythm, with the aim of identifying patients more suitable for this treatment. Specifically, these metrics include the mean and standard deviation of the relative wavelet energy ($RWEm$ and $RWEs$) across the different scales that encompass AA and an entropy-based metric that computes the stationary wavelet entropy variability ($SWEnV$). These metrics have previously demonstrated their capability to discern between SR and AF episodes~\cite{garcia2016application,rodenas2017combined}. Additionally, the predictors previously proposed by Hernandez et~al.~\cite{hernandez2014preoperative} ($fWP$, $DAF$, and $SampEn$), are also evaluated as reference parameters to test their predictive potential.
	

\section{Materials}
\textmk{Given the lack of public databases containing preoperative ECG recordings along with other clinical variables of permanent AF patients undergoing Maze and during their subsequent follow-up, a proprietary database was used in this study. It} was constructed from segments of 20 s in length, extracted from preoperative standard 12-lead ECGs recorded from 53 patients---24 men and 29 women with an average age of 68 $\pm$ 9 years. The rationale behind selecting segments of this duration was to strike a balance between capturing sufficient data for analysis, achieving good spectral resolution, and obtaining a stable estimate of the $DAF$. 

All the patients included in the study were diagnosed with permanent AF for a minimum of four months and received open-heart surgery in combination with a Cox--Maze IV procedure. The ECG signals were obtained with an amplitude resolution of 0.4~$\upmu$V and a sampling rate of 1000~Hz within two days prior to the surgical intervention. The high-resolution sampling allowed for capturing precise data, ensuring that the fine details of the ECG waveform were preserved for later analysis. After a one-year follow-up period, the data revealed that 23 out of the 53 patients (43.40\%) relapsed into AF, whereas the remaining 30 patients (56.60\%) had successfully maintained SR.

\section{Methods}

\begin{figure}[H]
    \includegraphics[width=0.99\textwidth]{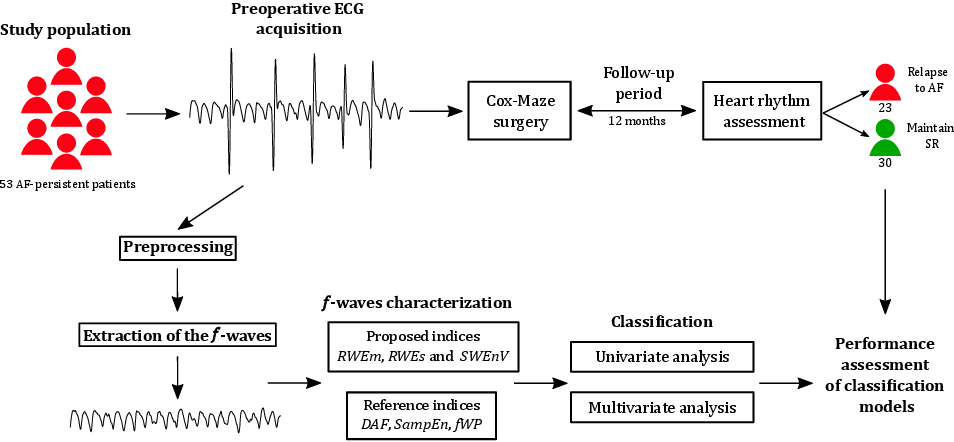}
    \caption{Overview of the study methodology, illustrating its various steps.}
    \label{fig:methodology_overview}
\end{figure}

The methods employed in this study were designed to offer a comprehensive understanding of the long-term risk assessment for AF recurrence following surgical ablation. Figure~\ref{fig:methodology_overview} provides an overview of the methodology employed in this study. The process encompasses data collection, signal preprocessing, feature extraction, predictive modeling, and outcome evaluation. This section outlines the details of each step in the methodology, providing a systematic explanation of the procedures and techniques employed to achieve the objectives of this study.

\subsection{Preprocessing}

\textmk{Lead V1 was selected for further analysis among the available leads due to previous research demonstrating better visualization of the AA in this signal~\cite{petrutiu2006atrial}. To facilitate its study, the signal was preprocessed. Thus, a filtering approach utilizing stationary wavelet transform (SWT) shrinking was initially employed to eliminate powerline interference in ECG recordings while preserving the signal's original morphology~\cite{garcia2018novel}. This approach has been extensively validated across a range of scenarios, including the analysis of ECG recordings obtained under both pathological and non-pathological conditions, as well as in the particular context of AF. Baseline wander was then removed through the utilization of an Infinite Impulse Response (IIR) high-pass filter with a 0.5~Hz cutoff frequency~\cite{dotsinsky2004optimization}. Additionally, an IIR low-pass filter with a cutoff frequency of 70~Hz was used to mitigate the high-frequency noise in the signal~\cite{sornmo2005biomedical}. Both filters were implemented utilizing a Chebyshev window, featuring a relative side-lobe attenuation of 40~dB}.

In order to efficiently detect the R-peaks in the preprocessed signal, a phasor transform-based approach was used~\cite{martinez2010application}. This method has been extensively validated on several databases manually annotated by experts, resulting in sensitivity and positive predictivity values exceeding 99.65\% and 99.70\%, respectively. In addition, it has exhibited proficiency in handling both normal and ectopic beats, a valuable feature within the context of AF.

Subsequently, for the computation of AF indices, an adaptive cancellation method was employed to eliminate ventricular activity and extract the AA signal~\cite{alcaraz2008adaptive}. This method involved aligning all QRS complexes with respect to their R-peaks. Cross-correlation was then employed to identify the 10 QRST complexes that exhibited the highest similarity to the one subject to cancellation. Principal component analysis was finally applied to derive a representative element from this set, which was further optimized to function as a cancellation template for the QRST under examination. This process resulted in an AA signal free from QRST contamination~\cite{alcaraz2008adaptive}.

\subsection{Proposed Method}

Once the AA signal was extracted from the preprocessed ECG, the aim of the subsequent step in the study methodology involved its characterization (Figure~\ref{fig:methodology_overview}). Prior investigations involving electrograms have unveiled a significant correlation between AA organization and the presence of wavelets circulating in the atria~\cite{barbaro2001mapping,faes2002method}. These studies have postulated that an increase in the number of wavelets is indicative of progressive electrical remodeling and a reduction in the effective refractory period of the atria, ultimately lowering the probability of AF reversion~\cite{everett2001frequency}. Furthermore, several studies focusing on the examination of surface ECG recordings have demonstrated that evaluating AA organization is a valuable method for predicting AF termination in a variety of situations, including the outcome of ECV~\cite{alcaraz2010review}, paroxysmal AF spontaneous termination~\cite{Chiarugi2007} or, last but not least, the anticipation of the AF outcome after Cox--Maze surgery at discharge~\cite{hernandez2014preoperative}. The last work combined the $DAF$, $SampEn$, and $fWP$ indices, obtaining promising predictive results~\cite{hernandez2014preoperative}.

Similarly, the analysis of the relative wavelet energy (RWE) and the stationary wavelet entropy (SWEn) variability in the AA signal has also revealed their capability to discern between SR and AF episodes when analyzed in the TQ interval of the ECG~\cite{garcia2016application,rodenas2017combined}. Therefore, the evaluation of these features could help characterize the variability in the morphological pattern of AA in AF rhythm with the aim of identifying patients who are more likely to maintain SR long term after Cox--Maze surgery.
 
The subsequent sections delve into the specific application of RWE and SWEn variability in this work and the evaluation of the $DAF$, $SampEn$, and $fWP$ as reference parameters. It is noteworthy that following preprocessing, the AA signal was divided into non-overlapped excerpts for the computation of the novel metrics used in this study. After experimenting with various sizes ranging from 0.6 s to 1.2 s at intervals of 0.1 s, the length of the excerpts used to derive the metrics explained below was established as 0.8 s.

\subsubsection{Relative Wavelet Energy}

The computation of the novel predictors proposed in this study entailed analyzing the AA signal using a method based on the wavelet transform (WT). This transformative technique allows for the decomposition of the signal into various time and frequency scales, emphasizing distinct signal properties and characteristics~\cite{mallat1999wavelet}. Certainly, the WT has established itself as a valuable tool for scrutinizing transients, aperiodicities, and other non-stationary signal features. It excels at accentuating subtle changes in signal morphology across the pertinent scales~\cite{mallat1999wavelet}. This algorithm, known for its non-redundant information, has found extensive application in various biomedical contexts~\cite{addison2005wavelet}. Its implementation is straightforward, involving a set of finite impulse response filters, including both low-pass and high-pass filters, followed by a decimation process~\cite{mallat1999wavelet}. It is pertinent to note that the mathematical intricacies of the WT are omitted from this manuscript (for in-depth details, readers are referred to~\cite{addison2005wavelet, mallat1999wavelet}).

However, it is imperative to highlight that the discrete WT faces issues of repeatability and robustness when handling short signals~\cite{asgari2015automatic, mallat1999wavelet}. Consequently, this study champions the utilization of the stationary wavelet transform (SWT). One key feature of the SWT is its time-invariance, ensuring that the number of wavelet coefficients at every level of decomposition is the same as the samples of the original signal~\cite{garcia2016application}.

Considering that the spectral content of the AA signal typically falls within the range of 3--12 Hz~\cite{sornmo2005biomedical} and given the 1 kHz sampling rate of the recordings, an eight-level wavelet decomposition was applied. In this way, the relative wavelet energy of AA will be mainly
concentrated in the sixth, seventh, and eighth scales. Notably, the SWT computation employed a sixth-order Daubechies wavelet function, consistent with previous works~\cite{garcia2016application}. Subsequently, we computed the RWE for the sixth, seventh, and eighth wavelet scales within each 0.8 s interval using the following equation

\begin{equation}\label{eq:RWE}
\mathrm{RWE}_j=\frac{\sum_{k=1}^PC(j,k)^2}{\sum_{j=1}^N\sum_{k=1}^PC(j,k)^2},
\end{equation}

\noindent where $C(j,k)$ is the series of wavelet coefficients of scale $j$ with translation $k$, $N$ is the number of wavelet decomposition levels, and $P$ is the length of $C(j,k)$ \cite{asgari2015automatic}. Finally, the mean and standard deviation values of the RWE for the sixth ($RWEm6$ and $RWEs6$), seventh ($RWEm7$ and $RWEs7$), and eighth scales ($RWEm8$ and $RWEs8$) were computed for each AA signal to evaluate its inter-segment variability.

\subsubsection{Stationary Wavelet Entropy Variability}
The second novel predictor introduced in this study involved estimating the variability of SWEn along the extracted AA signal. In terms of its operation, $SWEn$ was computed in each segment of the signal, resulting in the time series, $SWEn(n)$, where $n$ represents the number of intervals. This entropy-based metric measures the morphological complexity of a waveform by breaking it down into various time-frequency scales and subsequently calculating Shannon entropy on the relative energy distributions within those scales~\cite{garcia2016application}, i.e.,

\begin{equation}\label{eq:SWEn}
\mathrm{SWEn(n)}=-\sum_{j=1}^N \mathrm{RWE_j(n)}\cdot  \log(\mathrm{RWE_j(n)}).
\end{equation}

\textmk{Then, SWEn variability ($SWEnV$), a nonlinear index previously used to measure time-series regularity~\cite{rodenas2017combined}, was utilized to quantify the inter-segment variability in the morphological complexity of AA. Justifying the application of this nonlinear index to AF is grounded in the presence of nonlinearity at the cellular level in the diseased heart~\cite{bollmann2000quantification}. The temporal variability of the fibrillatory waves ($f$-waves) reflected on the AA signal, assessed through SWEn, was examined using an algorithm based on $SampEn$. This algorithm, commonly used to estimate RR interval-series regularity, measures the recurrence of similar patterns and was computed as described in~\cite{Lake2011}, i.e., }

\begin{equation} \label{eq:COSEn}
\begin{split}
SWEnV(SWEn, m, r, n) &= \left( SampEn(SWEn, m, r, n) + \ln(2r)\right) - \ln(\overline{SWEn(n)}) \\
&= (SampEn(SWEn, m, r, n) + \ln(2r)) - \ln \left(\frac{1}{n}\sum_{l=1}^n SWEn(l)\right),
\end{split}
\end{equation}

\noindent with $m$ being set to 1 sample, $r$ to 15\% tolerance, and $n$ to 25 intervals, as recommended in previous works~\cite{rodenas2017combined}. This metric provides low values for highly organized signals, whereas high values are associated with more disorganized waveforms~\cite{Lake2011}.

\subsection{Reference Methods}
For comparison purposes, the three indexes introduced by the only previous work that predicted AF recurrence at discharge following a Cox--Maze intervention were also studied~\cite{hernandez2014preoperative}. They were the $DAF$, $SampEn$, and $fWP$, and were computed as follows~\cite{hernandez2014preoperative}. 

Taking into account previous studies that have demonstrated a dominant frequency component in the AA signal during AF, typically falling within the range of 3–9~Hz~\cite{sornmo2005biomedical}, the $DAF$ was determined as the frequency with the highest amplitude within the specified range in the AA power spectrum. For this purpose, the power spectral density of AA was estimated using a Welch periodogram. The process involved the use of a Hamming window with a length of 4096 points, ensuring a 50\% overlap between adjacent windowed sections, and implementing an 8192-point fast Fourier transform~\cite{welch1967use}. $SampEn$ was defined as the negative natural logarithm of the conditional probability that two sequences, similar for $m$ points, will continue to be similar at point $m + 1$, excluding self-matches from the probability calculation~\cite{richman2000physiological,hernandez2014preoperative}. The $fWP$ represented the energy conveyed by the $f$-waves within the analyzed AA interval~\cite{hernandez2014preoperative} and was considered a strong indicator of the amplitude of the AA signal~\cite{alcaraz2009time}. This index was determined through the computation of the root-mean-square value of the AA segment~\cite{alcaraz2009time}. To prevent potential influences from the ECG amplitude, such as variations in recording gain factors, electrode impedance, and skin conductivity, this metric was normalized as a percentage of the R-peak magnitude~\cite{hernandez2014preoperative}.

\subsection{Classification Performance Analysis}

This section delves into a comprehensive evaluation of the individual prognostic parameters, aiming to uncover their effectiveness in predicting SR maintenance twelve months following the Cox--Maze procedure. For this purpose, a 5-fold cross-validation approach was considered to ensure robust results. This approach entailed dividing the data into 5 equally sized folds and conducting a 5-times repeated training-validation process. In each iteration, a different fold was chosen for validation, while the model was trained on the remaining folds~\cite{Refaeilzadeh2009}. Moreover, the database division was stratified to guarantee that each fold served as a representative sample of the entire dataset. A decision tree algorithm with a maximum number of 5 splits was considered to train the prediction model in each iteration, giving priority to simplicity in the interpretation of the forecasting methodology.

Although this validation process provided valuable insights into the model's ability to generalize beyond the specific dataset used for development, the validation of each single variable was repeated 100 times, reshuffling the data in each cycle in order to reduce the bias resulting from a single division of the data into 5 folds~\cite{Refaeilzadeh2009}. The classification results obtained by each 5-fold cross-validation procedure were summarized through the receiver operating characteristic (ROC) curve. This tool generates a plot illustrating the true positive rate or sensitivity (Se) compared to the false positive rate (i.e., 1-specificity) at different threshold settings for the scores provided by the prediction models. Sensitivity (Se) quantifies the ratio of correctly identified patients who will be in SR after twelve months, and specificity (Sp) reveals the effectiveness of the model in identifying patients who will remain in AF. Moreover, Acc assesses the overall probability of obtaining correct predictions by considering both true positives and true negatives in the context of the total patient population. The threshold chosen for the optimal separation of both groups of patients was that providing the highest Acc. The area under the ROC curve (AUC) was also calculated to provide a comprehensive assessment of the classification performance, independent of any specific threshold~\cite{Habibzadeh2016}. Furthermore, determining the positive predictive value (PPV) and negative predictive value (NPV) offered insights into the proportions of positive and negative samples correctly identified as true positives and true negatives, respectively. Finally, the values of these metrics were averaged for the 100 cycles conducted.
 
Furthermore, multivariate analysis was also performed to explore complementary information between the single features included in the study, with the aim of improving the prediction of the Cox--Maze outcome. Before that, a forward sequential feature selection technique was used to automatically choose the optimal combination of features for the multivariate model. This technique sequentially adds features to an empty candidate subset until a particular stopping condition is satisfied. A 5-fold cross-validation method was performed in each step to fit a model, \textmk{which was based on a decision tree classifier like the previous univariate models;} train it; and return a loss measure, i.e., the misclassification rate. This algorithm minimizes that loss measure until adding more features to the classification model does not help to achieve it~\cite{R_ckstie__2011}. The described method was repeated 50 times in order to mitigate inherent bias in the data partition and acquire the most representative subset of features. In this way, the most repeated combinations of single features were selected to build several decision tree-based multivariate classification models. These models were evaluated similarly to the single features, i.e., by running 100 5-fold cross-validation loops and obtaining the averaged Se, Sp, Acc, AUC, PPV, and NPV. \textmk{Additionally, the three single parameters used for comparison purposes---the $DAF$, $SampEn$, and $fWP$---were also combined via a decision tree classifier and validated in the same way to serve as a reference.}


\section{Results}

The obtained classification outcomes are presented in Table~\ref{tab:results_individual}, including the performance metrics for the individual predictors introduced in this study (i.e., $RWEm6$, $RWEs6$, $RWEm7$, $RWEs7$, $RWEm8$, $RWEs8$, and $SWEnV$) and those examined for comparison purposes (i.e., $DAF$, $SampEn$, and $fWP$). 

\begin{table}[H]\small
\renewcommand\arraystretch{1.38}
\caption{{Classification} performance results of individual parameters in predicting SR maintenance or relapse into AF 12 months after the Cox--Maze procedure.}
\label{tab:results_individual}
\begin{tabularx}{\textwidth}{L{12mm}CCCCCC}
\toprule
\textbf{Feature}  & \textbf{Se (\%)} & \textbf{Sp (\%)} & \textbf{Acc (\%)} & \textbf{AUC (\%)} & \textbf{PPV (\%)} & \textbf{NPV (\%)} \\
\noalign{\hrule height 0.5pt}
\rowcolor{white} 
$RWEm6$ & 58.47 & 69.61 & 63.31 & 58.13 & 71.51 & 56.24 \\
\rowcolor{gray!20} 
$RWEs6$ & 83.20 & 27.73 & 59.13 & 56.54 & 60.03 & 55.86 \\
\rowcolor{white} 
$RWEm7$ & 91.24 & 23.85 & 61.99 & 54.79 & 60.98 & 67.61 \\
\rowcolor{gray!20} 
$RWEs7$ & 80.83 & 67.09 & 74.87 & 71.25 & 76.21 & 72.85 \\
\rowcolor{white} 
$RWEm8$ &  89.30 & 28.09 & 62.74 & 58.96 & 61.83 & 66.82 \\
\rowcolor{gray!20} 
$RWEs8$ & 70.91 & 51.53 & 62.50 & 59.38 & 65.62 & 57.59 \\
\rowcolor{white} 
$SWEnV$ &  88.21 & 41.79 & 68.07 & 55.98 & 66.41 & 73.10 \\
\rowcolor{gray!20}
$DAF$ &  87.93 & 28.66 & 62.21 & 60.00 & 61.65 & 64.54 \\
 \rowcolor{white} 
$SampEn$ & 83.58 & 40.94 & 65.07 & 58.40 & 64.86 & 65.65 \\
\rowcolor{gray!20} 
$fWP$ &  86.33 & 29.36 & 61.61 & 59.00 & 61.45 & 62.22 \\
\noalign{\hrule height 1.0pt}
\end{tabularx}
\end{table}

As can be observed, most of the individual predictors exhibited limited predictive capabilities. However, an encouraging finding emerged, with four novel parameters ($RWEm6$, $RWEs7$, $RWEs8$, and $SWEnV$) demonstrating PPVs exceeding 65\%, although only two of them  ($SWEnV$ and $RWEs7$) demonstrated NPVs exceeding 70\%. $SWEnV$ yielded a classification Acc of 68.07\%, whereas within the features based on the RWE, $RWEs7$ stood out as the most promising individual predictor, with Se and Sp values of 80.83\% and 67.09\%, respectively, and an Acc of almost 75\%.

\begin{figure}[H]
	\includegraphics[width=1\textwidth]{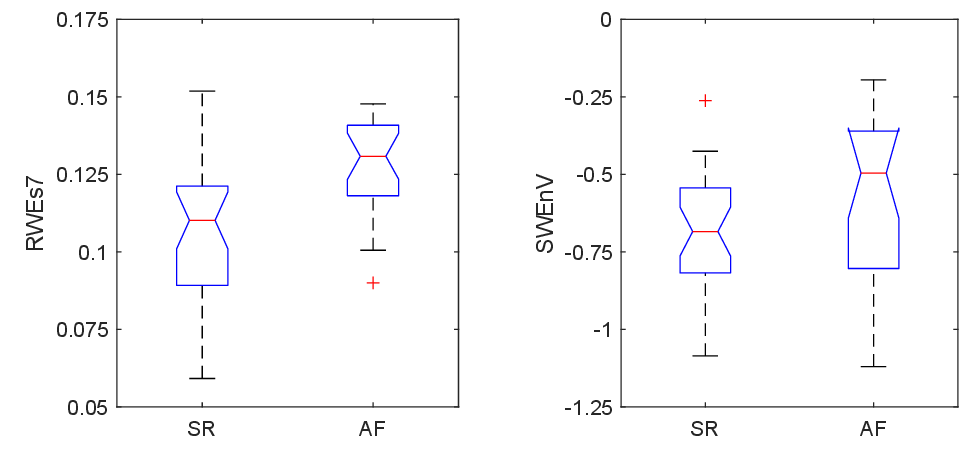}
	\caption{{Box-plot} 
 distribution of the standard deviation of the relative wavelet energy at the 7th scale, $RWEs7$, and the stationary wavelet entropy variability, $SWEnV$, obtained for patients enrolled in the study who maintained sinus rhythm (SR) and those who relapsed into atrial fibrillation (AF) one year after the Cox--Maze procedure.}
	\label{fig:boxplot}
\end{figure}

The distribution of values generated by these two most relevant features for both groups of patients, i.e., those who maintained SR and those who relapsed into AF one year after Cox--Maze surgery, is reflected in the box plot diagrams in Figure~\ref{fig:boxplot}. Of note is that the group of patients who experienced a relapse into AF presented higher values of $RWEs7$ and $SWEnV$ compared to the group of patients for whom the Cox--Maze treatment \mbox{was successful.}

In pursuit of enhancing predictive performance beyond what the single parameters could achieve, we employed a strategic multivariate approach aimed at maintaining a straightforward decision tree algorithm with a maximum number of five splits. The forward sequential feature selection technique provided the combinations of single features presented in Table~\ref{tab:results_pairs}. As can be seen, the multivariate models combined no more than three features to achieve the minimum classification error, which is the objective of the proposed feature selection algorithm. Moreover, it was also verified that adding more features to these subsets did not improve the classification results shown in this table.

\begin{table}[H]\small
\renewcommand\arraystretch{1.4}
\caption{\hl{Classification} 
 performance results of multivariate analysis in predicting SR maintenance or relapse into AF 12 months after the Cox--Maze procedure.}
\label{tab:results_pairs}
\begin{tabularx}{\textwidth}{L{43mm}CCCCCC}
\toprule
\textbf{Features Used in the Model} & \textbf{Se~(\%)} & \textbf{Sp~(\%)} & \textbf{Acc~(\%)} & \textbf{AUC~(\%)} & \textbf{PPV~(\%)} & \textbf{NPV~(\%)} \\
\noalign{\hrule height 0.5pt}
\rowcolor{white} 
$RWEs7$ and $fWP$ & 80.33 & 67.70 & 74.85 & 71.33 & 76.44 & 72.52 \\
\rowcolor{gray!20} 
$RWEs7$ and $SWEnV$ &  80.30 & 79.22 & 79.83 & 77.31 & 83.44 & 75.51 \\
\rowcolor{white} 
$RWEs7$, $SWEnV$, and $fWP$   &  80.93 & 79.35 & 80.25 & 77.73 & 83.64 & 76.14 \\
\rowcolor{gray!20} 
$RWEs6$, $RWEs7$, and $SWEnV$ & 78.87 & 76.87 & 78.00 & 74.50 & 81.64 & 73.61 \\
\rowcolor{white} 
$RWEm6$, $RWEs7$, and $SWEnV$ & 80.03 & 74.78 & 77.75 & 74.76 & 80.54 & 74.17 \\
\rowcolor{gray!20} 
$RWEs7$, $RWEs8$, and $SWEnV$ & 79.43 & 73.43 & 76.83 & 72.77 & 79.59 & 73.24 \\
\rowcolor{white} 
$RWEs7$, $SWEnV$, and $SampEn$ & 75.50 & 71.26 & 73.66 & 71.44 & 77.41 & 69.04 \\
\noalign{\hrule height 1.0pt}
\end{tabularx}
\end{table}

Notably, the combinations of two or three predictors consistently demonstrated improved predictive power compared to the individual parameters. Among the combinations formed by pairs of features, a decision tree using $SWEnV$ and $RWEs7$ emerged as the most successful. This model achieved the highest levels of prediction performance and accuracy, boasting an Acc of 79,83\%. Furthermore, it provided highly balanced values of Se and Sp of 80.30\% and 79.22\%, respectively, in comparison with the individual parameters. This innovative approach not only surpassed the predictive capabilities of the individual parameters but also offered a powerful yet simple method to predict long-term SR maintenance after the Cox--Maze procedure. 

Regarding the combinations of three indices, only the one composed of $RWEs7$, $SWEnV$, and $fWP$ managed to slightly outperform the previously highlighted predictor pair. Specifically, adding the $fWP$ feature to this pair increased the classification Acc by an absolute value of 0.42\%, becoming the only combination that exceeded 80\%. \textmk{It should be noted that all the models combining two or three automatically selected variables exhibited between 10 and 15\% better classification in terms of all the performance metrics compared to the model composed of the three reference indices $DAF$, $SampEn$, and $fWP$, which reported Se, Sp, Acc, AUC, PPV, and NPV values of 61.13\%, 60.61\%, 60.91\%, 64.32\%, 66.93\%, and 54.45\%, respectively.}


\section{Discussion}

Surgical ablation through the Cox--Maze procedure is strongly recommended when treating AF in patients undergoing cardiac surgery~\cite{hindricks2021}. However, the Cox--Maze intervention presents a limited success rate of approximately 65\% for permanent AF patients \cite{Sharples2018}. In fact, current AF management guidelines highlight the importance of the proper selection of optimal candidates and their tailored subsequent management, especially for preventing and minimizing the impact of AF recurrence after the intervention~\cite{hindricks2021}. In this context, preoperatively assessing the likelihood of AF recurrence after the Cox--Maze procedure serves strategic purposes by enabling healthcare professionals to plan personalized AF treatments. Using this approach in clinical decision making minimizes unnecessary interventions, the associated treatment costs, and the risks for patients with a lower probability of maintaining SR, and avoids potentially aggressive pharmacological treatments and their associated side effects for patients more suitable to undergoing the Cox--Maze procedure.

\textmk{Every indirect comparison of classification results obtained through different experimental frameworks should be considered with caution. Indeed, this kind of comparison might not be always fair because classification results can be highly variable as a function of the analyzed databases and the validation strategy used (e.g., resubstitution validation, hold-out validation, or cross-validation, among others). Nonetheless, they still enable outlining the main research lines in the state of the art. Table~\ref{tab:comparison} summarizes the most relevant previous works dealing with preoperative predictions of the Cox--Maze surgery. As can be observed,} most of these studies focused on the clinical and anatomical predictors associated with atrial function, achieving classification Acc values of about 75\%. For instance, Chen and colleagues~\cite{chen2004preoperative} identified the preoperative left atrial size as an independent predictor of successful SR maintenance in patients with permanent AF and mitral valve disease. In fact, they reported Se and Sp values of 50\% and 86.2\% when classification was conducted through linear discriminant analysis. This model provided a cutoff value of 56.25 cm$^2$ in the atrial area, associating smaller values with a higher likelihood of a favorable outcome when considering a mean follow-up period of 38 months.

\begin{table}[H]\small
	\caption{\textmk{Main features and classification results achieved by the most relevant previous works dealing with the prediction of AF recurrence after Cox--Maze surgery.}}\label{tab:comparison}
	\label{tab:resumen_estudios}
	
\begin{adjustwidth}{-\extralength}{0cm}
\begin{tabular}{m{2.5cm}<{\raggedright}m{2cm}<{\raggedright}m{5cm}<{\raggedright}m{3.8cm}<{\raggedright}m{3.1cm}<{\raggedright}}
		\toprule
		\textbf{Study}  & \textbf{Kind of AF} & \textbf{Relevant Single Features} & \textbf{Classification Model} & \textbf{Best Results} \\
		\midrule
		Chen et~al.~\cite{chen2004preoperative}  & Permanent & Left atrial area & Linear discriminant analysis with resubstitution validation & Se = 50.0\%; Sp = 86.2\% \\
        \midrule
		Wu et~al.~\cite{wu2017}    & Persistent & AF duration\par Left atrial diameter\par Right atrial area\par Intake of beta-blockers  & Logistic regression with \par resubstitution validation & Se = 79.9\%; Sp = 73.3\% Acc = 74.9\%  \\
        \midrule
		Cao et~al.~\cite{Cao2017}    & Persistent  & AF duration\par B-type natriuretic peptide\par Heart rate \par Left atrial diameter & Logistic regression with \par resubstitution validation & Se = 75.1\%; Sp = 81.5\% Acc = 75.8\%  \\

		\bottomrule
	\end{tabular}
\end{adjustwidth}
\end{table}

\begin{table}[H]\ContinuedFloat
\small
\caption{{\em Cont.}}

\begin{adjustwidth}{-\extralength}{0cm}
\begin{tabular}{m{2.5cm}<{\raggedright}m{2cm}<{\raggedright}m{5cm}<{\raggedright}m{3.8cm}<{\raggedright}m{3.1cm}<{\raggedright}}
		\toprule
		\textbf{Study}  & \textbf{Kind of AF} & \textbf{Relevant Single Features} & \textbf{Classification Model} & \textbf{Best Results} \\

        \midrule
		Jiang et~al.~\cite{jiang2023machine}  & Paroxysmal and \par persistent & AF duration\par Left ventricular ejection fraction\par Neutrophil--lymphocyte ratio\par Left atrial diameter\par Heart rate\par Rhythm after surgery, & Extreme gradient boosting with 5-fold cross-validation & Se = 63.3\%; \mbox{Acc = 80.2\%}; \mbox{AUC = 76.8\%}  \\
        \midrule
		Kakuta et~al.~\cite{kakuta2021risk}  & Persistent & $f$-wave voltage in V1\par AF duration\par Left atrial volume index\par Age & Logistic regression \par with hold-out validation & AUC = 78.0\%  \\
        \midrule
		This work            & Permanent & $SWEnV$\par $RWEs7$\par $fWP$ & Decision tree with \mbox{100 repetitions} of 5-fold cross-validation  &Se = 80.9\%; \hl{Sp~=~79.4\%;} Acc=80.3\%; \mbox{AUC = 77.7\%}  \\
		\bottomrule
	\end{tabular}
\end{adjustwidth}
\end{table}

A study by Wu and colleagues~\cite{wu2017} also assessed AF recurrence after the Cox--Maze procedure with concomitant mitral surgery through different clinical and anatomical features. They concluded that a longer persistent AF duration (>59.5 months) and a larger preoperative left atrial diameter (>59.85 mm) were predictors for negative long-term outcomes, with Acc values of approximately 75\%. Cao and colleagues~\cite{Cao2017} also evaluated the predictive capacity of the AF duration and left atrial diameter. Although their results showed the same trends as those in Wu et al.'s work~\cite{wu2017}, the Acc values of these predictors decreased to 55.3\% and 68.7\%, respectively. However, the study revealed better performance for the heart rate estimated before the intervention, with an Acc of 75.8\%. These discrepancies between studies highlight that there is no strong evidence to consider anatomical features as reliable single predictors of AF recurrence after Cox--Maze surgery.

To shed more light on this aspect, in a recent work by Jiang and colleagues~\cite{jiang2023machine}, the authors explored multiple machine learning models, combining 58 clinical variables to predict AF recurrence 5 years after the Cox--Maze procedure and valve surgery. The proposed models exhibited AUC values between 73.20\% and 76.80\%, revealing that the most significant preoperative features were the AF duration, left ventricular ejection fraction, neutrophil--lymphocyte ratio, left atrial diameter, and heart rate.

\textmk{In contrast to these previous works, the present study has addressed for the first time AF recurrence prediction after Cox--Maze surgery by analyzing ECG-based features focused on quantifying several properties of the AA signal. To the best of our knowledge,} only Hernández and colleagues have conducted a similar analysis, but in their case, the aim was to preoperatively anticipate Cox--Maze outcomes at discharge~\cite{hernandez2014preoperative}. This information could be helpful in the optimization of preoperative drug therapy planning and the anticipation of ECV-related decisions after the intervention~\cite{hernandez2014preoperative}, but it is not sufficient to achieve the aforementioned benefits of predicting AF recurrence in the mid-term regarding the selection of optimal candidates for the surgery and tailored scheduling of their subsequent follow-up. Moreover, Hernández et al. only analyzed the well-known parameters $DAF$, $SampEn$, and $fWP$, whereas the present work explored a wider set of novel features to characterize the variability in the morphological pattern of AA over time. Specifically, these metrics included the mean and standard deviation of the RWE ($RWEm$ and $RWEs$) across different scales encompassing the AA signal and the entropy-based metric $SWEnV$, which have previously demonstrated their capability to discern between SR and AF episodes~\cite{garcia2016application,rodenas2017combined}. 

When considered in isolation, none of these novel metrics achieved extremely high predictive rates of AF recurrence after Cox--Maze surgery, as shown in Table~\ref{tab:results_individual}. However, it is worth noting that $RWEs7$ stood out as the most promising individual predictor, reporting Se, Sp, and Acc values of 80.93\%, 67.09\%, and almost 75\%, respectively. This outcome aligns with expectations since the seventh-scale $RWE$ covers the most relevant frequency range of the $f$-waves (i.e., 3--12 Hz) when the ECG signal is sampled at 1~kHz. In terms of classification, this parameter was followed by the variable $SWEnV$, with an Acc value of 68.07\%). However, $SWEnV$ showed a significant Se--Sp imbalance, also observed in the other parameters, with an Se of 88.21\% and an Sp of 41.79\%. The classification performance of the remaining indices ($RWEm6$, $RWEs6$, $RWEm7$, $RWEm8$, and $RWEs8$) in terms of Acc and AUC values was similar to that of the previous indices considered for comparison purposes, i.e., $DAF$, $SampEn$, and $fWP$. In fact, all these indices reported Acc and AUC values of about 60--62\% and lower than 60\%, respectively.  

Nonetheless, the classification results were significantly enhanced when the information provided by the proposed predictors was combined through a multivariate approach. In this respect, all the automatically selected combinations of single parameters consistently outperformed their classification results, with significant improvements in terms of Acc and the Se--Sp balance observed, as reflected in Table~\ref{tab:results_pairs}.  Specifically, only the combination of the two best single predictors, $SWEnV$ and $RWEs7$, produced the most impressive outcomes, with Se and Sp values of 80.30\% and 79.22\%, respectively. This model excelled in predicting AF recurrence one year after the Cox--Maze procedure, reaching a classification Acc of 79.83\%. This outcome, along with the simplicity in the model's interpretation offered by the use of a decision tree classifier, indicates that the combination of these two predictors is particularly simple and effective in forecasting one-year postoperative Cox--Maze success.

Regarding the easy interpretation of this model, it should also be remarked that the data distribution generated by the two combined parameters aligns with observations from previous studies~\cite{rodenas2017combined}, which have established a connection between increased irregularity in the morphological complexity of AA in AF rhythms compared to SRs and higher values of the $SWEnV$. Moreover, a positive correlation between the degree of AA organization and the probability of maintaining SR after different AF treatments, including catheter ablation, has also been previously noted~\cite{Takahashi:2006aa}. Therefore, the higher values of the $SWEnV$ observed for the patients who relapsed into AF compared to those who maintained SR during the follow-up (see Figure~\ref{fig:boxplot}) reflected consistent behavior. On the other hand, although the RWE has demonstrated high effectiveness as a reliable discriminator for distinguishing AF from other rhythms~\cite{garcia2016application}, its performance diminished when challenged to discern subtle variations among AF patients. This performance decline can be attributed to the variability introduced when applied to AF signals, primarily stemming from the presence of harmonics in the $DAF$, causing energy redistribution across different scales. To address this challenge, the study introduced the standard deviation of the RWE, denoted as $RWEs$, which emerged as a robust discriminator for AF recurrence when evaluated on the seventh scale, denoted as $RWEs7$. This feature exhibited higher values in patients who relapsed into AF after the follow-up period of the Cox--Maze procedure (see Figure~\ref{fig:boxplot}), which was associated with higher variability in the AA signals. 

The conducted forward sequential feature selection also provided combinations of three single predictors, but in this case, any model improved the results when combining the $SWEnV$ and $RWEs7$ indices, except when these two variables were complemented by the $fWP$ metric. This suggests that taking into account the time-domain information of the $f$-waves can also help obtain a more complete picture of AA organization for the predictive model of the Cox--Maze outcome. However, only a slight and non-significant improvement of less than 0.5\% with respect to the two-feature model was observed in this case. \textmk{It is worth noting that this classification model, with Acc and AUC values of about 80\% and 78\%, respectively, exhibited comparable or slightly better performance compared to previous works, even when using a more unbiased and robust validation methodology, such as 100 repetitions in a 5 fold cross-validation approach~\cite{Refaeilzadeh2009} (see Table~\ref{tab:comparison}). Although some aspects related to datasets, like the size, heterogeneity, and follow-up period, might impact the classification results of predictive models, it is still possible to highlight that the proposed ECG-based analysis of the $f$-wave variability over time can provide useful and complementary information for the previously introduced clinical and anatomical predictors of AF recurrence after Cox--Maze surgery~\cite{chen2004preoperative, wu2017, Cao2017, jiang2023machine}. } 

This idea has also been suggested by Kakuta and coworkers~\cite{kakuta2021risk}, who recently proposed a 10-point risk score model for preoperative predictions of Cryo--Maze success in the long term by integrating clinical and anatomical information with an $f$-wave characteristic measured from the surface ECG. Indeed, as risk factors, the model considered the $f$-wave voltage in the lead V1 of preoperative 12-lead ECG recordings <0.2 mV (4 points), the preoperative AF duration > 5 years (3 points), the left atrial volume index > 100 mL/m$^2$ (2 points), and an age > 70 years (1 point). An accumulated score > 7 points could predict high rates of AF recurrence with an AUC of 78\%, considering a maximum prediction time horizon of approximately 5 years after the procedure. Hence, these results demonstrate that the integration of clinical information and ECG-derived variables can provide a more comprehensive understanding and enhance the prediction accuracy of SR maintenance after Cox--Maze surgery. Therefore, future research endeavors should evaluate the predictive performance of the novel $RWEs7$ and $SWEnV$ metrics from preoperative ECG recordings together with the clinical and anatomical information of AF patients.

As a final remark, it is essential to recognize certain limitations in the methodology developed. Firstly, the results may be subject to slight variations due to the relatively low number of patients included in the analysis. A more comprehensive understanding of the predictive capabilities may be attained by expanding the dataset to include a larger number of recordings. Furthermore, the proposed analysis of the AA signal was exclusively focused on lead V1. While lead V1 is widely acknowledged for its superior visibility and discrimination of AA from ventricular activity, it should be noted that other ECG leads may contain valuable information that has not been unexplored in this study. Finally, another limitation of this work is that neither clinical nor anatomical features were incorporated into the predictive model. This omission will be addressed in future investigations aimed at combining ECG signal analysis features with clinical information to provide deeper insights into the atrial function and arrhythmia mechanism, thereby enhancing our understanding of Cox--Maze surgery outcome predictions.


\section{Conclusions}

A novel methodology was introduced in this study for predicting long-term Cox--Maze procedure outcomes in patients with permanent AF. Emphasis was placed on the significance of preoperative information from the ECG in characterizing the variability in the morphological pattern of AA in AF rhythm with the aim of identifying patients more suitable for this treatment. When evaluated in isolation, the individual predictors were found to lack sufficient predictive power for AF recurrence. However, the combination of the stationary wavelet entropy variability, $SWEnV$, and the standard deviation of the relative wavelet energy at the seventh scale, $RWEs7$, consistently outperformed the individual predictors. A decision tree classifier was used to build the prediction models, giving priority to simplicity and easy clinical interpretation of the results. The specialization of the model in identifying patients at risk of mid-term AF recurrence offers a significant step toward personalized care and tailored procedures. This emphasis on a tailored approach to AF management is instrumental for clinicians and researchers seeking to enhance post-surgical care and outcomes.

\vspace{6pt}


\authorcontributions{Conceptualization, P.E., J.R., M.G., F.H., J.M.G.-B., R.A., and J.J.R.; methodology, P.E., J.R., M.G., R.A., and J.J.R.; software, P.E., J.R., and M.G.; validation, P.E., J.R., M.G., F.H., R.A., and J.J.R.; resources, J.M.G.-B., F.H., and J.J.R.; data curation, P.E., J.R., and M.G.; writing---original draft preparation, P.E., J.R., and M.G.; writing---review and editing, P.E., J.R., M.G., F.H., J.M.G.-B., R.A., and J.J.R. All authors have read and agreed to the published version of the manuscript.}

\funding{\textls[-25]{\hl{This research} 
 has received financial support from public grants PID2021-123804OB-I00, PID2021-00X128525-IV0, and TED2021-130935B-I00 of the Spanish Government, 10.13039/501100011033,} in conjunction with the European Regional Development Fund (EU), SBPLY/21/180501/000186, from Junta de Comunidades de Castilla-La Mancha, and AICO/2021/286 from Generalitat Valenciana. Pilar Escribano holds the 2020-PREDUCLM-15540 scholarship co-financed by the European Social Fund (ESF) operating program 2014–2020 of Castilla-La Mancha.}


\institutionalreview{This study was conducted according to the guidelines of the Declaration of Helsinki, complied with Law 14/2007, of~3 July, on~Biomedical Research and other Spanish regulations, and was approved by the Ethical Review Board of the General University Hospital Consortium of Valencia (Valencia, Spain) with protocol code 153/2020.}

\informedconsent{Written informed consent was granted from all the subjects participating in the present research. All acquired data were anonymized before processing.}

\dataavailability{The data supporting the reported results presented in this study are available on request from the corresponding author.} 


\conflictsofinterest{The authors declare that there are no conflicts of interest related to this study. They have no financial or personal associations with commercial entities that could be perceived as having an interest in the area of the submitted manuscript. The study's funders played no role in shaping the study's design, data collection, analysis, or data interpretation. They were not involved in the manuscript's writing, nor did they participate in the decision to publish the study's results.}

\abbreviations{List of Acronyms}{

\noindent
\begin{tabular}{@{}ll}
	
	\hl{Acronym} 
			&	Definition \\
	{AF} & Atrial fibrillation\\	
	SR           	& Sinus rhythm\\	
	ECV          	& Electrical cardioversion\\
	CA				& Catheter ablation\\		
	AA				& Atrial activity\\
	ECG				& Electrocardiogram\\	
	$fWP$			& Amplitude of the fibrillatory waves\\
	$DAF$			& Dominant atrial frequency\\
	$SampEn$		& Sample entropy\\
	$RWEm$			& Mean of the relative wavelet energy\\
	$RWEs$			& Standard deviation of the relative wavelet energy\\
	$SWEnV$			& Stationary wavelet entropy variability\\
	WT				& Wavelet transform\\
	SWT				& Stationary wavelet transform\\
	IIR				& Infinite impulse response\\
	$f$-waves		& Fibrillatory waves\\	
	ROC				& Receiver operating characteristic\\
	Se				& Sensitivity\\
	Sp				& Specificity\\
	Acc				& Accuracy\\
	AUC				& Area under the ROC curve\\
	PPV				& Positive predictive value\\
	NPV				& Negative predictive value\\

\end{tabular}
}

\begin{adjustwidth}{-\extralength}{0cm}

\reftitle{\highlighting{References} 
}


\begin{thebibliography}{999}

\bibitem[Lippi et~al.(2021)Lippi, Sanchis-Gomar, and Cervellin]{Lippi:2021aa}
Lippi, G.; Sanchis-Gomar, F.; Cervellin, G.
\newblock Global epidemiology of atrial fibrillation: An increasing epidemic
  and public health challenge.
\newblock {\em Int. J. Stroke} {\bf 2021}, {\em 16},~217--221.
\newblock {\url{https://doi.org/10.1177/1747493019897870}}.

\bibitem[Zoni-Berisso et~al.(2014)Zoni-Berisso, Lercari, Carazza, and
  Domenicucci]{zoni2013epidemiology}
Zoni-Berisso, M.; Lercari, F.; Carazza, T.; Domenicucci, S.
\newblock Epidemiology of atrial fibrillation: European perspective.
\newblock {\em Clin. Epidemiol.} {\bf 2014}, \emph{6},  213--220.
\newblock {\url{https://doi.org/10.2147/CLEP.S47385}}.

\bibitem[Nations et~al.(2017)Nations, of~Economic, and
  Social~Affairs]{unitednations2017world}
United Nations and Department of Economic and Social Affairs, Population Division.
\newblock World Population Ageing 2017---Highlights (ST/ESA/SER. A/397),  \hl{2017.} \hl{Available online:}
\newblock {\url{https://www.un.org/en/development/desa/population/publications/pdf/ageing/WPA2017_Highlights.pdf}} \hl{(accessed on 12 September 2023).}

\bibitem[Hindricks et~al.(2021)Hindricks, Potpara, Dagres, Arbelo, Bax,
  Blomstr{\"o}m-Lundqvist, Boriani, Castella, Dan, Dilaveris,
  et~al.]{hindricks2021}
Hindricks, G.; Potpara, T.; Dagres, N.; Arbelo, E.; Bax, J.J.;
  Blomstr{\"o}m-Lundqvist, C.; Boriani, G.; Castella, M.; Dan, G.A.; Dilaveris,
  P.E.;  et~al.
\newblock 2020 ESC Guidelines for the diagnosis and management of atrial
  fibrillation developed in collaboration with the European Association for
  Cardio-Thoracic Surgery (EACTS) The Task Force for the diagnosis and
  management of atrial fibrillation of the European Society of Cardiology (ESC)
  Developed with the special contribution of the European Heart Rhythm
  Association (EHRA) of the ESC.
\newblock {\em Eur. Heart J.} {\bf 2021}, {\em 42},~373--498.
\newblock {\url{https://doi.org/10.1093/eurheartj/ehaa612}}.

\bibitem[Di~Carlo et~al.(2020)Di~Carlo, Zaninelli, Mori, Consoli, Bellino,
  Baldereschi, Sgherzi, Gradia, D'Alfonso, Cattarinussi, and et~al.]{Carlo2020}
Di~Carlo, A.; Zaninelli, A.; Mori, F.; Consoli, D.; Bellino, L.; Baldereschi,
  M.; Sgherzi, B.; Gradia, C.; D'Alfonso, M.G.; Cattarinussi, A.;  et~al.
\newblock Prevalence of atrial fibrillation subtypes in Italy and projections
  to 2060 for Italy and Europe.
\newblock {\em J. Am. Geriatr. Soc.} {\bf 2020}, {\em
  68},~2534–2541.
\newblock {\url{https://doi.org/10.1111/jgs.16748}}.

\bibitem[Morin et~al.(2016)Morin, Bernard, Madias, Rogers, Thihalolipavan, and
  Estes~III]{morin2016state}
Morin, D.P.; Bernard, M.L.; Madias, C.; Rogers, P.A.; Thihalolipavan, S.;
  Estes~III, N.M.
\newblock The state of the art: Atrial fibrillation epidemiology, prevention,
  and treatment.
\newblock  \emph{Mayo Clin. Proc.}   \textbf{2016},
  \emph{91},  1778--1810.
\newblock {\url{https://doi.org/10.1016/j.mayocp.2016.08.022}}.

\bibitem[Deshmukh et~al.(2022)Deshmukh, Iglesias, Khanna, and
  Beaulieu]{deshmukh2022}
Deshmukh, A.; Iglesias, M.; Khanna, R.; Beaulieu, T.
\newblock Healthcare utilization and costs associated with a diagnosis of
  incident atrial fibrillation.
\newblock {\em Heart Rhythm O$_{2}$} {\bf 2022}, {\em 3},~577–586.
\newblock {\url{https://doi.org/10.1016/j.hroo.2022.07.010}}.

\bibitem[Carlsson et~al.(2003)Carlsson, Miketic, Windeler, Cuneo, Haun, Micus,
  Walter, Tebbe, and investigators]{carlsson2003randomized}
Carlsson, J.; Miketic, S.; Windeler, J.; Cuneo, A.; Haun, S.; Micus, S.;
  Walter, S.; Tebbe, U.; investigators, S.
\newblock Randomized trial of rate-control versus rhythm-control in persistent
  atrial fibrillation: The Strategies of Treatment of Atrial Fibrillation
  (STAF) study.
\newblock {\em J. Am. Coll. Cardiol.} {\bf 2003}, {\em
  41},~1690--1696.
\newblock {\url{https://doi.org/10.1016/S0735-1097(03)00332-2}}.

\bibitem[Van~Wagoner et~al.(2015)Van~Wagoner, Piccini, Albert, Anderson,
  Benjamin, Brundel, Califf, Calkins, Chen, Chiamvimonvat,
  et~al.]{vanwagoner2015progress}
Van~Wagoner, D.R.; Piccini, J.P.; Albert, C.M.; Anderson, M.E.; Benjamin, E.J.;
  Brundel, B.; Califf, R.M.; Calkins, H.; Chen, P.S.; Chiamvimonvat, N.;
  et~al.
\newblock Progress toward the prevention and treatment of atrial fibrillation:
  A summary of the Heart Rhythm Society Research Forum on the Treatment and
  Prevention of Atrial Fibrillation, Washington, DC, December 9--10, 2013.
\newblock {\em Heart Rhythm} {\bf 2015}, {\em 12},~e5--e29.
\newblock {\url{https://doi.org/10.1016/j.hrthm.2014.11.011}}.

\bibitem[Mont and Guasch(2017)]{mont2017atrial}
Mont, L.; Guasch, E.
\newblock Atrial fibrillation progression: How sick is the atrium?
\newblock {\em Heart Rhythm} {\bf 2017}, {\em 14},~808--809.
\newblock {\url{https://doi.org/10.1016/j.hrthm.2017.02.027}}.

\bibitem[Kik et~al.(2017)Kik, Mouws, Bogers, and de~Groot]{kik2017intra}
Kik, C.; Mouws, E.M.; Bogers, A.J.; de~Groot, N.M.S.
\newblock Intra-operative mapping of the atria: The first step towards
  individualization of atrial fibrillation therapy?
\newblock {\em Expert Rev. Cardiovasc. Ther.} {\bf 2017}, {\em
  15},~537--545.
\newblock {\url{https://doi.org/10.1080/14779072.2017.1340156}}.

\bibitem[Anselmino et~al.(2016)Anselmino, Matta, Castagno, Giustetto, and
  Gaita]{anselmino2016catheter}
Anselmino, M.; Matta, M.; Castagno, D.; Giustetto, C.; Gaita, F.
\newblock Catheter ablation of atrial fibrillation in chronic heart failure:
  state-of-the-art and future perspectives.
\newblock {\em Europace} {\bf 2016}, {\em 18},~638--647.
\newblock {\url{https://doi.org/10.1093/europace/euv368}}.

\bibitem[Schmidt et~al.(2020)Schmidt, Brugada, Arbelo, Laroche, Bayramova,
  Bertini, Letsas, Pison, and et~al.]{Schmidt:2020aa}
Schmidt, B.; Brugada, J.; Arbelo, E.; Laroche, C.; Bayramova, S.; Bertini, M.; Letsas, K.P.; Pison, L.; Romanov, A.; Scherr, D.; et al.
\newblock {Ablation strategies for different types of atrial fibrillation in
  Europe: Results of the ESC-EORP EHRA Atrial Fibrillation Ablation Long-Term
  registry}.
\newblock {\em Europace} {\bf 2020}, {\em 22},~558--566.
\newblock   {\url{https://doi.org/10.1093/europace/euz318}}.

\bibitem[Dretzke et~al.(2020)Dretzke, Chuchu, Agarwal, Herd, Chua, Fabritz,
  Bayliss, Kotecha, Deeks, Kirchhof, and Takwoingi]{Dretzke2020}
Dretzke, J.; Chuchu, N.; Agarwal, R.; Herd, C.; Chua, W.; Fabritz, L.; Bayliss,
  S.; Kotecha, D.; Deeks, J.J.; Kirchhof, P.;  et~al.
\newblock Predicting recurrent atrial fibrillation after catheter ablation: A
  systematic review of prognostic models.
\newblock {\em Europace} {\bf 2020}, {\em 22},~748--760.
\newblock {\url{https://doi.org/10.1093/europace/euaa041}}.

\bibitem[Sharples et~al.(2018)Sharples, Everett, Singh, Mills, Spyt, Abu-Omar,
  Fynn, Thorpe, Stoneman, Goddard, and et~al.]{Sharples2018}
Sharples, L.; Everett, C.; Singh, J.; Mills, C.; Spyt, T.; Abu-Omar, Y.; Fynn,
  S.; Thorpe, B.; Stoneman, V.; Goddard, H.;  et~al.
\newblock Amaze: A double-blind, multicentre randomised controlled trial to
  investigate the clinical effectiveness and cost-effectiveness of adding an
  ablation device-based maze procedure as an adjunct to routine cardiac surgery
  for patients with pre-existing atria.
\newblock {\em Health Technol. Assess.} {\bf 2018}, {\em 22},~1–132.
\newblock {\url{https://doi.org/10.3310/hta22190}}.

\bibitem[Cox(2014)]{cox2014brief}
Cox, J.L.
\newblock A brief overview of surgery for atrial fibrillation.
\newblock {\em Ann. Cardiothorac. Surg.} {\bf 2014}, {\em 3},~80--88.
\newblock {\url{https://doi.org/10.3978/j.issn.2225-319X.2014.01.05}}.

\bibitem[Damiano~Jr et~al.(2011)Damiano~Jr, Schwartz, Bailey, Maniar, Munfakh,
  Moon, and Schuessler]{damiano2011cox}
Damiano, R.J., Jr.; Schwartz, F.H.; Bailey, M.S.; Maniar, H.S.; Munfakh, N.A.;
  Moon, M.R.; Schuessler, R.B.
\newblock \textls[-5]{The Cox maze IV procedure: Predictors of late recurrence.}
\newblock {\em  J. Thorac. Cardiovasc. Surg.} {\bf 2011},
  {\em 141},~113--121.
\newblock {\url{https://doi.org/10.1016/j.jtcvs.2010.08.067}}.

\bibitem[Prasad et~al.(2003)Prasad, Maniar, Camillo, Schuessler, Boineau,
  Sundt~III, Cox, and Damiano~Jr]{prasad2003cox}
Prasad, S.M.; Maniar, H.S.; Camillo, C.J.; Schuessler, R.B.; Boineau, J.P.;
  Sundt, T.M., III; Cox, J.L.; Damiano, R.J., Jr.
\newblock The Cox maze III procedure for atrial fibrillation: Long-term
  efficacy in patients undergoing lone versus concomitant procedures.
\newblock {\em  J. Thorac. Cardiovasc. Surg.} {\bf 2003},
  {\em 126},~1822--1827.
\newblock {\url{https://doi.org/10.1016/S0022-5223(03)01287-X}}.

\bibitem[Henn et~al.(2015)Henn, Lancaster, Miller, Sinn, Schuessler, Moon,
  Melby, Maniar, and Damiano~Jr]{henn2015late}
Henn, M.C.; Lancaster, T.S.; Miller, J.R.; Sinn, L.A.; Schuessler, R.B.; Moon,
  M.R.; Melby, S.J.; Maniar, H.S.; Damiano, R.J., Jr.
\newblock Late outcomes after the Cox maze IV procedure for atrial
  fibrillation.
\newblock {\em  J. Thorac. Cardiovasc. Surg.} {\bf 2015},
  {\em 150},~1168--1178.
\newblock {\url{https://doi.org/10.1016/j.jtcvs.2015.07.102}}.

\bibitem[Engelsgaard et~al.(2018)Engelsgaard, Pedersen, Riber, Pallesen, and
  Brandes]{engelsgaard2018long}
Engelsgaard, C.S.; Pedersen, K.B.; Riber, L.P.; Pallesen, P.A.; Brandes, A.
\newblock The long-term efficacy of concomitant maze IV surgery in patients
  with atrial fibrillation.
\newblock {\em IJC Heart  Vasc.} {\bf 2018}, {\em 19},~20--26.
\newblock {\url{https://doi.org/10.1016/j.ijcha.2018.03.009}}.

\bibitem[Chen et~al.(2004)Chen, Chang, and Chang]{chen2004preoperative}
Chen, M.C.; Chang, J.P.; Chang, H.W.
\newblock Preoperative atrial size predicts the success of radiofrequency maze
  procedure for permanent atrial fibrillation in patients undergoing
  concomitant valvular surgery.
\newblock {\em Chest} {\bf 2004}, {\em 125},~2129--2134.
\newblock {\url{https://doi.org/10.1378/chest.125.6.2129}}.

\bibitem[Wu et~al.(2017)Wu, Chang, Chen, Cheng, and Chung]{wu2017}
Wu, C.C.; Chang, J.P.; Chen, M.C.; Cheng, C.I.; Chung, W.J.
\newblock Long-term results of radiofrequency maze procedure for persistent
  atrial fibrillation with concomitant mitral surgery.
\newblock {\em J. Thorac. Dis.} {\bf 2017}, {\em 9},~5176–5183.
\newblock {\url{https://doi.org/10.21037/jtd.2017.11.112}}.

\bibitem[Cao et~al.(2017)Cao, Xue, Zhou, Yu, Tang, and Wang]{Cao2017}
Cao, H.; Xue, Y.; Zhou, Q.; Yu, M.; Tang, C.; Wang, D.
\newblock Late outcome of surgical radiofrequency ablation for persistent
  valvular atrial fibrillation in China: A single-center study.
\newblock {\em J. Cardiothorac. Surg.} {\bf 2017}, {\em 12},  63.
\newblock {\url{https://doi.org/10.1186/s13019-017-0627-z}}.

\bibitem[Jiang et~al.(2023)Jiang, Song, Liang, Zhang, Tan, Sun, Guo, and
  Liu]{jiang2023machine}
Jiang, Z.; Song, L.; Liang, C.; Zhang, H.; Tan, H.; Sun, Y.; Guo, R.; Liu, L.
\newblock Machine learning-based analysis of risk factors for atrial
  fibrillation recurrence after Cox-Maze {IV} procedure in patients with atrial
  fibrillation and chronic valvular disease: A retrospective cohort study with
  a control group.
\newblock {\em Front. Cardiovasc. Med.} {\bf 2023}, {\em 10}, 1140670.
\newblock {\url{https://doi.org/10.3389/fcvm.2023.1140670}}.

\bibitem[Alcaraz and Rieta(2010)]{alcaraz2010review}
Alcaraz, R.; Rieta, J.J.
\newblock A review on sample entropy applications for the non-invasive analysis
  of atrial fibrillation electrocardiograms.
\newblock {\em Biomed. Signal Process.  Control} {\bf 2010}, {\em
  5},~1--14.
\newblock {\url{https://doi.org/10.1016/j.bspc.2009.11.001}}.

\bibitem[Chiarugi et~al.(2007)Chiarugi, Varanini, Cantini, Conforti, and
  Vrouchos]{Chiarugi2007}
Chiarugi, F.; Varanini, M.; Cantini, F.; Conforti, F.; Vrouchos, G.
\newblock Noninvasive ECG as a tool for predicting termination of paroxysmal
  atrial fibrillation.
\newblock {\em IEEE Trans. Biomed. Eng.} {\bf 2007}, {\em
  54},~1399–1406.
\newblock {\url{https://doi.org/10.1109/tbme.2007.890741}}.

\bibitem[Hernandez et~al.(2014)Hernandez, Alcaraz, Hornero, and
  Rieta]{hernandez2014preoperative}
Hernandez, A.; Alcaraz, R.; Hornero, F.; Rieta, J.J.
\newblock Preoperative study of the surface ECG for the prognosis of atrial
  fibrillation maze surgery outcome at discharge.
\newblock {\em Physiol. Meas.} {\bf 2014}, {\em 35},~1409.
\newblock {\url{https://doi.org/10.1088/0967-3334/35/7/1409}}.

\bibitem[Garc{\'\i}a et~al.(2016)Garc{\'\i}a, R{\'o}denas, Alcaraz, and
  Rieta]{garcia2016application}
Garc{\'\i}a, M.; R{\'o}denas, J.; Alcaraz, R.; Rieta, J.J.
\newblock Application of the relative wavelet energy to heart rate independent
  detection of atrial fibrillation.
\newblock {\em Comput. Methods Programs Biomed.} {\bf 2016}, {\em
  131},~157--168.
\newblock {\url{https://doi.org/10.1016/j.cmpb.2016.04.009}}.

\bibitem[R\'{o}denas et~al.(2017)R\'{o}denas, Garc\'{i}a, Alcaraz, and
  Rieta]{rodenas2017combined}
R\'{o}denas, J.; Garc\'{i}a, M.; Alcaraz, R.; Rieta, J.J.
\newblock Combined nonlinear analysis of atrial and ventricular series for
  automated screening of atrial fibrillation.
\newblock {\em Complexity} {\bf 2017}, \emph{2017}, 2163610.
\newblock {\url{https://doi.org/10.1155/2017/2163610}}.

\bibitem[Petrutiu et~al.(2006)Petrutiu, Ng, Nijm, Al-Angari, Swiryn, and
  Sahakian]{petrutiu2006atrial}
Petrutiu, S.; Ng, J.; Nijm, G.M.; Al-Angari, H.; Swiryn, S.; Sahakian, A.V.
\newblock Atrial fibrillation and waveform characterization. A time domain
  perspective in the surface {ECG}.
\newblock {\em IEEE Eng. Med. Biol. Mag.} {\bf 2006},
  {\em 25},~24--30.
\newblock {\url{https://doi.org/10.1109/emb-m.2006.250505}}.

\bibitem[Garc{\'\i}a et~al.(2018)Garc{\'\i}a, Mart{\'\i}nez-Iniesta,
  R{\'o}denas, Rieta, and Alcaraz]{garcia2018novel}
Garc{\'\i}a, M.; Mart{\'\i}nez-Iniesta, M.; R{\'o}denas, J.; Rieta, J.J.;
  Alcaraz, R.
\newblock A novel wavelet-based filtering strategy to remove powerline
  interference from electrocardiograms with atrial fibrillation.
\newblock {\em Physiol. Meas.} {\bf 2018}, {\em 39},~115006.
\newblock {\url{https://doi.org/10.1088/1361-6579/aae8b1}}.

\bibitem[Dotsinsky and Stoyanov(2004)]{dotsinsky2004optimization}
Dotsinsky, I.; Stoyanov, T.
\newblock Optimization of bi-directional digital filtering for drift
  suppression in electrocardiogram signals.
\newblock {\em J. Med Eng.   Technol.} {\bf 2004}, {\em
  28},~178--180.
\newblock {\url{https://doi.org/10.1080/03091900410001675996}}.

\bibitem[S{\"o}rnmo and Laguna(2005)]{sornmo2005biomedical}
S{\"o}rnmo, L.; Laguna, P.
\newblock {\em Bioelectrical Signal Processing in Cardiac and Neurological
  Applications}; Academic Press: Cambridge, MA, USA,  2005.

\bibitem[Mart{\'\i}nez et~al.(2010)Mart{\'\i}nez, Alcaraz, and
  Rieta]{martinez2010application}
Mart{\'\i}nez, A.; Alcaraz, R.; Rieta, J.J.
\newblock Application of the phasor transform for automatic delineation of
  single-lead {ECG} fiducial points.
\newblock {\em Physiol. Meas.} {\bf 2010}, {\em 31},~1467.
\newblock {\url{https://doi.org/10.1088/0967-3334/31/11/005}}.

\bibitem[Alcaraz and Rieta(2008)]{alcaraz2008adaptive}
Alcaraz, R.; Rieta, J.J.
\newblock Adaptive singular value cancelation of ventricular activity in
  single-lead atrial fibrillation electrocardiograms.
\newblock {\em Physiol. Meas.} {\bf 2008}, {\em 29},~1351--1369.
\newblock {\url{https://doi.org/10.1088/0967-3334/29/12/001}}.

\bibitem[Barbaro et~al.(2001)Barbaro, Bartolini, Calcagnini, Censi, Morelli,
  and Michelucci]{barbaro2001mapping}
Barbaro, V.; Bartolini, P.; Calcagnini, G.; Censi, F.; Morelli, S.; Michelucci,
  A.
\newblock Mapping the organization of atrial fibrillation with basket catheters
  part I: Validation of a real-time algorithm.
\newblock {\em Pacing Clin. Electrophysiol.} {\bf 2001}, {\em
  24},~1082--1088.
\newblock {\url{https://doi.org/10.1046/j.1460-9592.2001.01082.x}}.

\bibitem[Faes et~al.(2002)Faes, Nollo, Antolini, Gaita, and
  Ravelli]{faes2002method}
Faes, L.; Nollo, G.; Antolini, R.; Gaita, F.; Ravelli, F.
\newblock A method for quantifying atrial fibrillation organization based on
  wave-morphology similarity.
\newblock {\em IEEE Trans. Biomed. Eng.} {\bf 2002}, {\em
  49},~1504--1513.
\newblock {\url{https://doi.org/10.1109/TBME.2002.805472}}.

\bibitem[Everett et~al.(2001)Everett, Kok, Vaughn, Moorman, and
  Haines]{everett2001frequency}
Everett, T.H.; Kok, L.C.; Vaughn, R.H.; Moorman, R.; Haines, D.E.
\newblock Frequency domain algorithm for quantifying atrial fibrillation
  organization to increase defibrillation efficacy.
\newblock {\em IEEE Trans. Biomed. Eng.} {\bf 2001}, {\em
  48},~969--978.
\newblock {\url{https://doi.org/10.1109/10.942586}}.

\bibitem[Mallat(1999)]{mallat1999wavelet}
Mallat, S.
\newblock {\em A Wavelet Tour of Signal Processing}; Elsevier: Berlin/Heidelberg, Germany, 1999.
\newblock {\url{https://doi.org/10.1016/B978-0-12-466606-1.X5000-4}}.

\bibitem[Addison(2005)]{addison2005wavelet}
Addison, P.S.
\newblock Wavelet transforms and the ECG: A review.
\newblock {\em Physiol. Meas.} {\bf 2005}, {\em 26},~R155.
\newblock {\url{https://doi.org/10.1088/0967-3334/26/5/R01}}.

\bibitem[Asgari et~al.(2015)Asgari, Mehrnia, and Moussavi]{asgari2015automatic}
Asgari, S.; Mehrnia, A.; Moussavi, M.
\newblock Automatic detection of atrial fibrillation using stationary wavelet
  transform and support vector machine.
\newblock {\em Comput. Biol. Med.} {\bf 2015}, {\em
  60},~132--142.
\newblock {\url{https://doi.org/10.1016/j.compbiomed.2015.03.005}}.

\bibitem[Bollmann(2000)]{bollmann2000quantification}
Bollmann, A.
\newblock Quantification of electrical remodeling in human atrial fibrillation.
\newblock {\em Cardiovasc. Res.} {\bf 2000}, {\em 47},~207--209.
\newblock {\url{https://doi.org/10.1016/S0008-6363(00)00133-4}}.

\bibitem[Lake and Moorman(2011)]{Lake2011}
Lake, D.E.; Moorman, J.R.
\newblock Accurate estimation of entropy in very short physiological time
  series: The problem of atrial fibrillation detection in implanted ventricular
  devices.
\newblock {\em Am. J. Physiol.-Heart  Circ. Physiol.}
  {\bf 2011}, {\em 300},~H319--H325.
\newblock {\url{https://doi.org/10.1152/ajpheart.00561.2010}}.

\bibitem[Welch(1967)]{welch1967use}
Welch, P.
\newblock The use of fast Fourier transform for the estimation of power
  spectra: A method based on time averaging over short, modified periodograms.
\newblock {\em IEEE Trans. Audio Electroacoust.} {\bf 1967},
  {\em 15},~70--73.
\newblock {\url{https://doi.org/10.1109/TAU.1967.1161901}}.

\bibitem[Richman and Moorman(2000)]{richman2000physiological}
Richman, J.S.; Moorman, J.R.
\newblock Physiological time-series analysis using approximate entropy and
  sample entropy.
\newblock {\em Am. J. Physiol.-Heart  Circ. Physiol.}
  {\bf 2000}, {\em 278},~H2039--H2049.
\newblock {\url{https://doi.org/10.1152/ajpheart.2000.278.6.H2039}}.

\bibitem[Alcaraz and Rieta(2009)]{alcaraz2009time}
Alcaraz, R.; Rieta, J.J.
\newblock Time and frequency recurrence analysis of persistent atrial
  fibrillation after electrical cardioversion.
\newblock {\em Physiol. Meas.} {\bf 2009}, {\em 30},~479.
\newblock {\url{https://doi.org/10.1088/0967-3334/30/5/005}}.

\bibitem[Refaeilzadeh et~al.(2009)Refaeilzadeh, Tang, and
  Liu]{Refaeilzadeh2009}
Refaeilzadeh, P.; Tang, L.; Liu, H.
\newblock {Cross-Validation}. In {\em Encyclopedia of Database Systems}; {Liu},
  L., Özsu, M.T., Eds.; Springer: New York, NY, USA,  2009; pp. 532--538.
\newblock {\url{https://doi.org/10.1007/978-0-387-39940-9_565}}.

\bibitem[Habibzadeh et~al.(2016)Habibzadeh, Habibzadeh, and
  Yadollahie]{Habibzadeh2016}
Habibzadeh, F.; Habibzadeh, P.; Yadollahie, M.
\newblock {On determining the most appropriate test cut-off value: The case of
  tests with continuous results}.
\newblock {\em Biochem. Med.} {\bf 2016}, {\em 26},~297--307.
\newblock {\url{https://doi.org/10.11613/BM.2016.034}}.

\bibitem[R{\"u}ckstie{\ss} et~al.(2011)R{\"u}ckstie{\ss}, Osendorfer, and
  van~der Smagt]{R_ckstie__2011}
R{\"u}ckstie{\ss}, T.; Osendorfer, C.; van~der Smagt, P.
\newblock {Sequential feature selection for classification}. In {\em {AI} 2011:
  Advances in Artificial Intelligence}; Springer: Berlin/Heidelberg, Germany, 2011; pp.
  132--141.
\newblock {\url{https://doi.org/10.1007/978-3-642-25832-9_14}}.

\bibitem[Kakuta et~al.(2021)Kakuta, Fukushima, Minami, Saito, Kawamoto,
  Tadokoro, Ikuta, Kobayashi, and Fujita]{kakuta2021risk}
Kakuta, T.; Fukushima, S.; Minami, K.; Saito, T.; Kawamoto, N.; Tadokoro, N.;
  Ikuta, A.; Kobayashi, J.; Fujita, T.
\newblock Novel risk score for predicting recurrence of atrial fibrillation
  after the Cryo-Maze procedure.
\newblock {\em Eur. J. Cardio-Thorac. Surg.} {\bf 2021}, {\em
  59},~1218--1225.
\newblock {\url{https://doi.org/10.1093/ejcts/ezaa468}}.

\bibitem[Takahashi et~al.(2006)Takahashi, Sanders, Ja{\"\i}s, Hocini, Dubois,
  Rotter, Rostock, and et~al.]{Takahashi:2006aa}
Takahashi, Y.; Sanders, P.; Jais, P.; Hocini, M.; Dubois, R.; Rotter, M.; Rostock, T.; Nalliah, C.J.; Sacher, F.; Clémenty, J.; et al.
\newblock Organization of frequency spectra of atrial fibrillation: Relevance
  to radiofrequency catheter ablation.
\newblock {\em J. Cardiovasc. Electrophysiol.} {\bf 2006}, {\em 17},~382--388.
\newblock {\url{https://doi.org/10.1111/j.1540-8167.2005.00414.x}}.

\end{thebibliography}

\PublishersNote{}
\end{adjustwidth}
\end{document}